\documentclass[
 reprint,
nofootinbib,
 amsmath,amssymb,
 aps,
 prl,
]{revtex4-2}

\usepackage{graphicx}
\usepackage{dcolumn}
\usepackage{bm}
\usepackage{hyperref}
\usepackage[utf8]{inputenc}
\usepackage{color}
\usepackage{ulem}

\newcommand{\red}[1]{{\color{red} #1}}

\newcommand{\abar}{a_{\text{bar}}}

\newcommand{\Mbar}{M_{\text{bar}}}

\begin{document}


\title{Presence of a Fundamental Acceleration Scale in Galaxy Clusters}

\date{\today}

\author{Douglas Edmonds}
\email{bde12@psu.edu}
\affiliation{%
Department of Physics, Penn State Hazleton, Hazleton, PA, 18202, U.S.A.
}%

\author{Djordje Minic}
\email{dminic@vt.edu}
\author{Tatsu Takeuchi}
\email{takeuchi@vt.edu}
\affiliation{%
Department of Physics, 
Virginia Tech, Blacksburg VA, 24061, U.S.A.
}%

\begin{abstract}
An acceleration scale of order $10^{-10}\mathrm{m/s^2}$
is implicit in the baryonic Tully-Fisher and baryonic Faber-Jackson relations,
independently of any theoretical preference or bias. 
We show that the existence of this scale in the baryonic Faber-Jackson relation is 
most apparent when data from pressure supported systems of vastly different
scales including globular clusters, elliptical galaxies, and galaxy clusters
are analyzed together.
This suggests the relevance of the acceleration scale $10^{-10}\mathrm{m/s^2}$ to
structure formation processes at many different length scales and could be pointing to a heretofore
unknown property of dark matter.

\end{abstract}

\maketitle


\section{\label{sec:intro}Introduction}

The observation of astronomical correlations 
has often lead to profound discoveries
that not only explained where those correlations came from, but also changed the
overall paradigm of physics and astronomy. 
The most well known example must be Kepler's three laws (1st and 2nd Laws 1609, 3rd Law 1619), 
which preceded the discoveries of Newton (Principia, 1687).
The Titius-Bode Law (Titius 1766, Bode 1772), which did lead to the discovery of Ceres (1801),  
but was not taken seriously since its subsequent failure to predict the correct orbital radius of Neptune (discovered 1846), 
may yet yield important insights connecting classical and quantum mechanics 
\cite{Blagg:1913,Scardigli:2005fr,Freund:2015nts,Aschwanden_2018}.
Early in the 20th century, 
Leavitt's Law (1908 \cite{Leavitt:1908vb,Leavitt:1912zz}) provided the first 
standard candle that enabled the determination of distances to
far away stars, while
the Hubble-Lema\^itre Law (1927 \cite{Lemaitre:1927zz}, 1929 \cite{Hubble:1929ig}) 
has fundamentally changed our view of the Universe.

In this letter,
we focus our attention on 
the baryonic versions of the Tully-Fisher (1977 \cite{Tully:1977fu}) 
and Faber-Jackson (1976 \cite{Faber:1976sn}) relations,
which are widely believed to encode important information on 
galaxy formation, and point out that an acceleration scale 
universal to both relations is implicit in the data,
independently of any theoretical preference or bias.
Furthermore, we argue that the baryonic Faber-Jackson relation can be extended to 
the Galaxy Cluster scale, suggesting that the said universal acceleration scale may be relevant
to structure formation processes of the Universe at many different length scales.
Given the deep connection between dark matter (DM) and structure formation,
this could be pointing to a heretofore unknown property of dark matter
not encompassed within the standard $\Lambda$CDM paradigm.

\section{The Baryonic Tully-Fisher Relation}

The baryonic Tully-Fisher relation (BTFR) \cite{Freeman:1999,McGaugh:2000sr} is an empirical relation 
observed in rotationally supported systems such as spiral galaxies,
which posits that the total baryonic mass of the system $M_\mathrm{bar}$
(the sum of the masses of the stars and gas)
is correlated with the asymptotic rotational velocity $v_\infty$ of 
objects orbiting the outskirts of the system as
\begin{equation}
M_\mathrm{bar}\;\propto\; v_\infty^n\;,
\label{BTFR}
\end{equation}
where $n\approx 4$ \cite{Aaronson_1983,Strauss:1995fz,Willick:1998yt,Bell:2000jt,Verheijen_2001,Kannappan_2004,Weiner_2006A,Weiner_2006B,Kassin_2007,Gurovich:2010jx,TorresFlores:2011uc,Zaritsky_2014,Iorio_2017}.
The original Tully-Fisher relation (TFR, 1977 \cite{Tully:1977fu})
was proposed as a relation between the 
absolute optical luminosity $L$ of the system and $v_\infty$ (determined from the width of the
21~cm \textsc{Hi} line) of the form
\begin{equation}
L \;\propto\; v_\infty^n\;.
\end{equation}
Such a relation would allow the measurement of $v_\infty$ to be used as a proxy for $L$,
and the comparison of $L$ with the apparent luminosity would then allow the
determination of the distance to the system.
Indeed, this method has been used to determine the Hubble parameter $H_0$
\cite{Tully:1977fu,Aaronson_1983,Madore_1998,Sakai_2000,Tutui_2001,Hendry_2001,Russell_2009,Schombert:2020pxm},
independently from type-1a supernovae \cite{Reid:2019tiq},
type-II supernovae \cite{deJaeger:2020zpb},
or CMB \cite{Ade:2015xua} observations.

The TFR and BTFR have been studied intensively by various groups (see Table~3 in \cite{TorresFlores:2011uc})
and the BTFR has been found to apply with smaller scatter than the original TFR 
to both HSB (high surface brightness) and LSB (low surface brightness) spiral galaxies,
as well as irregular \cite{TorresFlores:2011uc} and dwarf irregular galaxies \cite{Iorio_2017},
covering $\sim$5 decades of mass scale: $\Mbar = 10^{6\sim 11}M_\odot$. (See Fig.~23 of \cite{Iorio_2017}).

Note that the BTFR implies the universality of its slope
among these various rotationally supported systems (HSB, LSB, irregular, and dwarf irregular galaxies)
which for $n=4$ is
\begin{equation}
a_{\S} \;=\; \dfrac{v_\infty^4}{G\Mbar}\;.
\end{equation}
Here, $G$ is Newton's gravitational constant giving $a_{\S}$ the dimensions of an acceleration. 
The analysis of galactic rotation curve data in Ref.~\cite{McGaugh:2011ac} yields $a_{\S}=(1.6\pm 0.2)\,a_0$,
where $a_0\equiv 10^{-10}\,\mathrm{m/s^2}$.
We emphasize here that we are not invoking MOND \cite{Milgrom:1983ca,Milgrom:1983pn,Milgrom:1983zz} or any other theory to obtain this scale.
It is simply the slope of the BTFR.

A correlation between the total dark matter and total baryonic matter within a region surrounding a galaxy is to be
expected in the $\Lambda$CDM scenario, but given that galaxies are thought to have gone through various phases
including starbursts, emission of gasses, and multiple mergers during their evolutionary histories, 
a dynamic correlation involving a universal acceleration scale is surprising.
Though simple scaling arguments leading to the BTFR have been suggested \cite{Zwaan_1995,White_1997},
they do not necessarily predict $n=4$, or they rely on assumptions that are not satisfied by the data.

A recent $N$-body simulation of galaxy formation
from collapsing gas in a dark matter halo, which included modeled star formation feedback processes, reports
results which are consistent with $n=4$ BTFR \cite{SIMBA_2020}.
This could be an indication that the universality of $a_\S$
in spirals is already implicit in the standard $\Lambda$CDM dynamics
and no new physics is necessary to explain it. (See also \cite{Mo:2000pu,Kaplinghat:2001me,Navarro:2016bfs}.)
However, it remains to be seen how robust the result is when 
further details of baryonic physics is included in the simulation model,
and other processes such as galaxy mergers are considered.

\section{The Baryonic Faber-Jackson Relation}

The correlation which parallels the BTFR for pressure supported systems such as elliptical galaxies 
is the baryonic Faber-Jackson relation (BFJR)
\begin{equation}
\Mbar \;\propto\; \sigma^n\;,
\end{equation}
where $\sigma$ is the line-of-sight (los) velocity dispersion.
Again, it was originally proposed (FJR, 1976 \cite{Faber:1976sn}) as a proportionality between the absolute optical luminosity $L$ of elliptical galaxies and a power of $\sigma$
\begin{equation}
L \;\propto\; \sigma^n\;,
\end{equation}
with $n\approx 4$ \cite{Faber:1976sn,Kormendy:1982}. However,
subsequent analyses of elliptical galaxy data have found the
power $n$ to be anywhere from about 3 to 5, depending on the dataset and analysis method used \cite{Kormendy:1982},  
and as low as 2 for dwarf galaxies \cite{Cody:2009}.
The BFJR can also be fit to globular cluster data, but there, there is considerable scatter 
so the meaning of the fit by itself is unclear \cite{Nella-Courtois:1999}.

Though the value of $n$ is not well constrained when the BFJR is fit
separately to elliptical galaxies or globular clusters,
it was pointed out in Ref.~\cite{Farouki:1983ApJ} (1983) 
that if both data sets are analyzed together, 
they are seen to cluster along an $n=4$ BFJR line spanning $\sim$8 decades of mass scale:
$\Mbar = 10^{4\sim 12}M_\odot$. (See Fig.~4 of \cite{Farouki:1983ApJ}
and FIG.~3 of this letter.)
And just as in the BTFR, $n=4$ BFJR would immediately imply the existence of an acceleration scale
\begin{equation}
a_{\varnothing} \;=\; \dfrac{\sigma^4}{G \Mbar}\;,
\label{eqn:a_elliptical}
\end{equation}
which is universal among elliptical galaxies and globular clusters.
In Ref.~\cite{Farouki:1983ApJ}, its value was found to be $a_{\varnothing} = (1.5_{-0.7}^{+1.4})\,a_0$,
the same order of magnitude as $a_\S$.
Again, we emphasize here that neither MOND nor any other theory 
was invoked to obtain this scale.  It is simply the slope of the BFJR.

Given that elliptical galaxies and globular clusters are very different systems
in their sizes, dark matter content, evolutionary histories, etc., 
it is quite surprising that they would cluster along the same $n=4$ BJFR.
And it is also mysterious why the universal accelerations for rotationally supported systems $a_{\S}$,
and that for pressure supported systems $a_{\varnothing}$, are both of order $a_0$.

\section{\label{sec:fj}The Baryonic Faber-Jackson Relation in Galaxy Clusters}

The discussion so far suggests that an acceleration scale of order $a_0 = 10^{-10}\,\mathrm{m/s^2}$
is universal to both rotationally and pressure supported systems including and below
the galaxy scale. 
Let us now shift our attention to galaxy clusters
and investigate whether the BFJR holds at that scale.
Here, we utilize 
a sample of 
64 galaxy clusters from the HIghest X-ray FLUx Galaxy Cluster Sample (HIFLUGCS) \cite{Reiprich:2001zv}. 
This is an X-ray flux-limited 
($f_X(0.1 \sim 2.4~\text{keV})\geq 2\times 10^{-11}\text{ergs}\cdot\text{s}^{-1}\cdot\text{cm}^{-2}$) 
sample of X-ray selected galaxy clusters based on the ROSAT All-Sky Survey \cite{ROSAT}.

Gas mass estimates and the los velocity dispersions for the galaxy clusters are taken from Ref.~\cite{Zhang:2010qk}. 
Since the gas mass cannot be measured directly, we must rely on estimates based upon measurements of other cluster parameters. 
One way to estimate the (total) mass for a galaxy cluster is to infer it from velocity dispersions via the virial theorem. 
However, for the work presented here, it is imperative that the mass estimates be independent of the measurement of velocity dispersions or any assumptions about the dynamics. 
In Ref.~\cite{Zhang:2010qk} 
gas mass estimates are determined from the X-ray surface brightness in the soft X-ray band, 
while los velocity dispersions are determined from the motions of the member galaxies,
thus satisfying this important criterion.
The total baryonic mass $\Mbar$ is then estimated from the total gas mass $M_\mathrm{gas}$.

\begin{figure}
\includegraphics[width=\columnwidth]{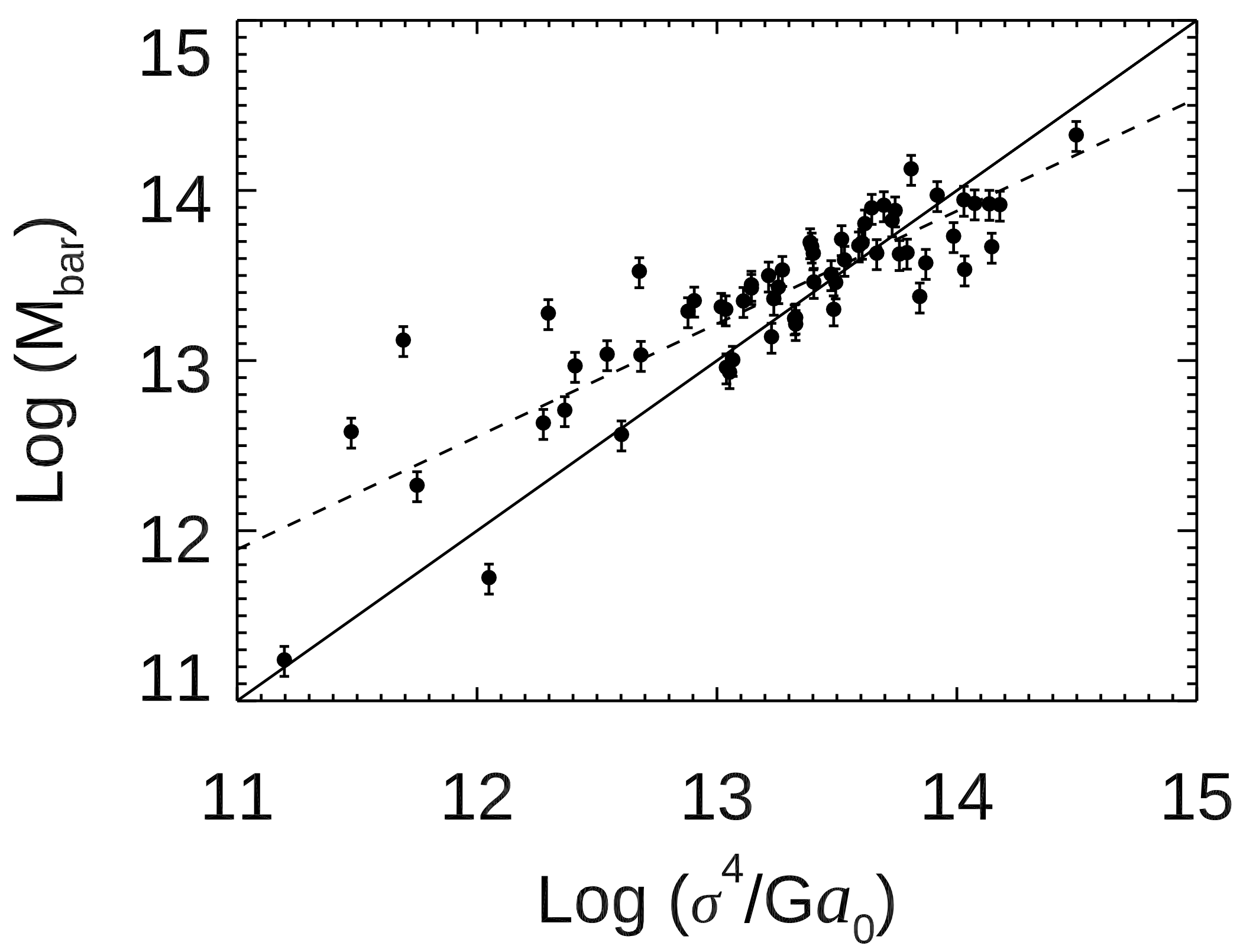}
\caption{A Faber-Jackson relation for galaxy clusters. The circles are galaxy cluster data with error bars, the dashed line is the best-fit to the data, and the solid line is locus of points where the baryonic mass is equal to the given function of velocity-dispersion.}
\label{fig:galclusters}
\end{figure}

In FIG.~\ref{fig:galclusters} we show a log-log plot of
$\Mbar/M_\odot$ versus $\sigma^4/GM_\odot a_0$ for the HIFLUGCS galaxy cluster set.
The solid black circles with error bars (statistical) are the data,
and a clear correlation between $\Mbar$ and $\sigma^4$ is manifest,
demonstrating the validity of the BFJR for galaxy clusters.
The solid black line with slope one is the $\Mbar=\sigma^4/G a_0$ line, which corresponds to $n=4$ BFJR with $a_{\varnothing} =a_0$,
while the dashed line indicates the line of best-fit to the data and corresponds to $n\approx 3$, 
consistent with the analysis of Ref.~\cite{Zhang:2010qk}.\footnote{%
Curiously, Ref.~\cite{Zhang:2010qk} finds $n\approx 4$ for the FJR.}
Thus, the galaxy cluster data prefers an $n\approx 3$ BFJR,
though we can see from inspection that the $n=4$ line is not a bad fit either,
especially when one notes that potentially large systematic errors are not shown.

\begin{figure}
\includegraphics[width=\columnwidth]{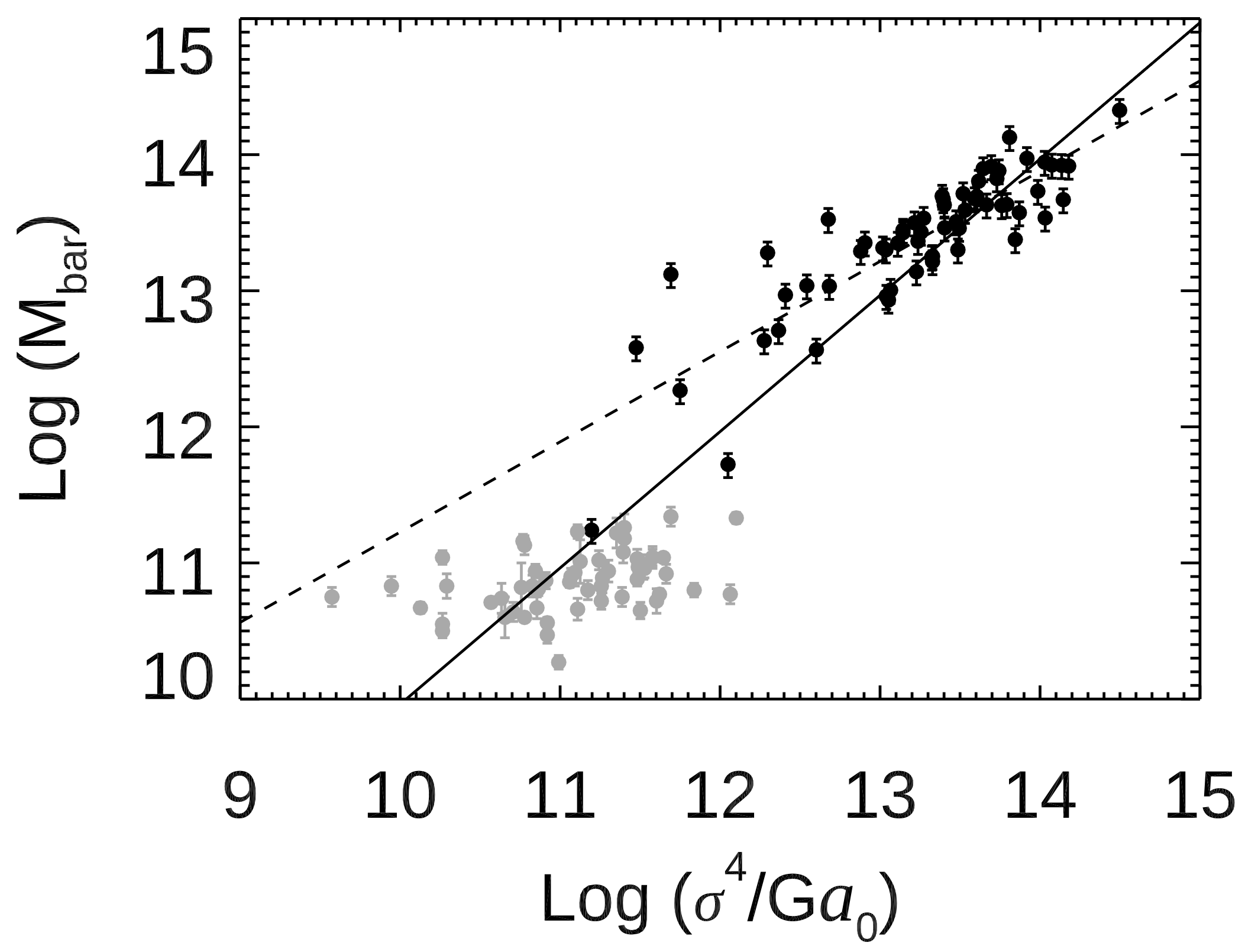}
\caption{The same as Figure 1, except we have included elliptical galaxies (grey circles), and the solid line is now the best-fit to the data set including both galaxy cluster and elliptical galaxies.}
\label{fig:ellipticals}
\end{figure}

Following Ref.~\cite{Farouki:1983ApJ}, let us include data from other pressure supported
structures.
In addition to the galaxy cluster data from FIG.~\ref{fig:galclusters} (black circles), 
FIG.~\ref{fig:ellipticals} shows baryonic mass estimates and los velocity dispersions of elliptical galaxies 
taken from Ref.~\cite{Belli_2014} (gray circles).
Fitting the BTFR to both data sets together, we obtain the solid black line
which corresponds to $n\approx 4$ and $a_{\varnothing}\approx 1.4\,a_0$.

\begin{figure}
\includegraphics[width=\columnwidth]{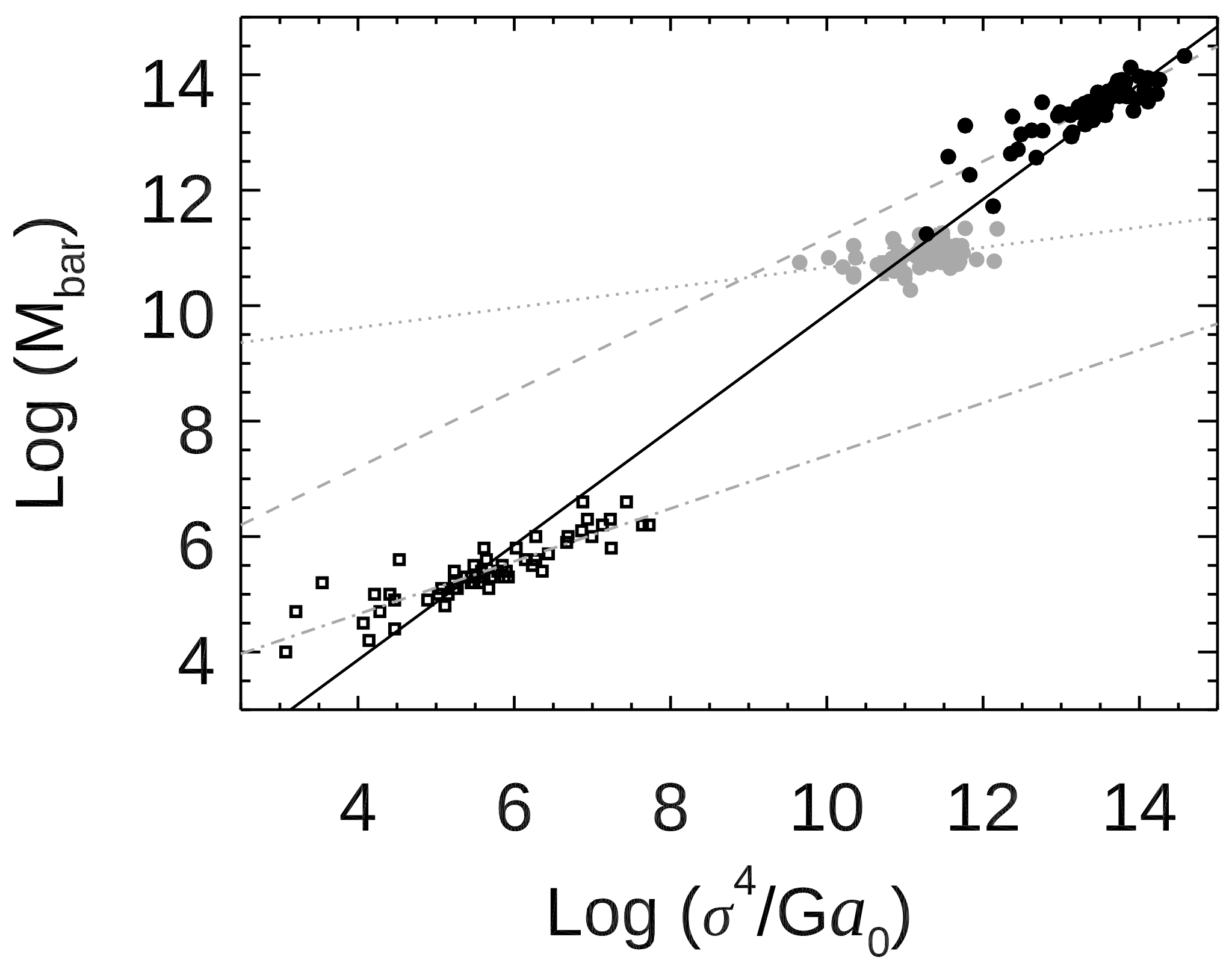}
\caption{The Faber-Jackson relation including galaxy clusters, elliptical galaxies, and globular clusters. The black and grey circles are the same as in Figures 1 and 2. The open squares represent data from globular clusters. The dashed line is the same as in Figure 1, the dotted and dot-dashed lines are the best-fit to the elliptical galaxies alone and the globular clusters alone, respectively. The solid line is the best-fit to all of the data taken together.}
\label{fig:globulars}
\end{figure}

Continuing on to smaller scales, FIG.~\ref{fig:globulars} adds 
baryonic mass estimates and los velocity dispersions for globular clusters taken from 
Ref~\cite{Meylan:1993yd,Meylan:1993yd2} (open squares) to the data already plotted in FIG.~\ref{fig:galclusters}.
While the mass estimates for globular clusters are not entirely independent of velocity dispersion measurements, they were based upon the King model of globular clusters (see \cite{King:1968} and references therein) for which dispersion is just one of the model parameters, rather than relying on observed correlations.
The dashed, dotted, and dot-dashed lines respectively indicate the best-fit
when the galaxy cluster, elliptical galaxy, and globular cluster data are fitted separately.
However, when the three data sets are fitted together, 
we obtain the black solid line which corresponds to $n\approx 4$ and $a_{\varnothing}\approx 1.3\,a_0$.
Thus, the $n=4$ BFJR is valid over $\sim$10 decades of mass scale $\Mbar = 10^{4\sim 14}M_\odot$,
and its slope is $a_{\varnothing} = O(a_0)$.

We note that the existence of the BFJR and $a_0$ in galaxy cluster data had been pointed out as early as 1994
by Sanders in Ref.~\cite{Sanders:1994}, in which the MONDian virial theorem, which includes $a_0$,
was used to relate $\sigma$ and $\Mbar$.
The analysis was extended to include globular clusters in Ref.~\cite{Sanders:2010aa}
to posit a ``universal Faber-Jackson relation'' similar to FIG.~\ref{fig:globulars}.
(See also Refs.~\cite{Durazo:2018jzn,Hernandez:2019xup,Bilek:2019}.)
Our current analysis demonstrates that the appearance of the BFJR straddling vastly different scales
does not depend on any dynamics, Newtonian or otherwise.
It is purely an observational fact.

Galaxy clusters are vastly different systems from galaxies in both scale and evolutionary histories. 
They typically consist of hundreds to thousands of galaxies, and the majority of their baryonic masses ($\sim$90\%) 
resides in the hot intergalactic gas.
Even if the dynamics of each member galaxy is governed by $a_\S$ or $a_\varnothing$,
why would the dynamics of galaxy clusters be governed by the same acceleration scale?
Is the information on $a_\S$ and $a_\varnothing$ transferred from the member galaxies to the cluster, or 
is the acceleration present in the structure formation dynamics independently of the scale?

\section{\label{sec:conclusions}Implications to Dark Matter and Structure Formation}

We have seen that an acceleration scale of order $a_0$ is universal to the BTF and BFJ relations
over many decades of scale covering many different types of structures.
This underscores the universal importance of $a_0$ in structure formation processes \cite{Edmonds:2020iji}.

There are several possibilities that may explain the presence of $a_0$. 
The simplest one is, of course, that it is emergent in the standard $\Lambda$CDM model \cite{Mo:2000pu,Kaplinghat:2001me}.
For spirals, there is some evidence in $N$-body simulations that this may indeed be the case 
\cite{Navarro:2016bfs,SIMBA_2020}.
We are not aware of similar results for ellipticals.
For galaxy clusters, the ongoing Cluster-EAGLE project \cite{Barnes:2017,Armitage:2017uee}
may provide us with clues.
We note, however, that the appearance of $a_0$ in simulations does not
immediately mean an understanding of why the same $a_0$ emerges over many decades of scale
in $\Lambda$CDM.

The second possibility is that $a_0$ is a fundamental parameter of nature,
which governs structure formation processes at all scales via some heretofore unknown physics.
It has been noted that $a_0 \approx cH_0/2\pi$ \cite{Milgrom:1986hz}, where $H_0$ is the Hubble parameter, 
suggesting that $a_0$ could be a proxy for $H_0$ or $\Lambda\sim 3H_0^2$.
Following the standard assumption that structure formation proceeds via gravity,
$a_0$ can be made to modify Einstein's equation,
\begin{equation}
G_{ab} +\Lambda g_{ab} = 8\pi G\,T_{ab}\;,
\label{Einstein}
\end{equation}
either by modifying gravity (left-hand side) or DM (right-hand side).

Models that modify DM to explain BTFR and BFJR have been proposed, 
for instance, in Refs.~\cite{Khoury:2014tka,Ho:2010ca, Ho:2011xc, Ho:2012ar}:
\cite{Khoury:2014tka} proposes a superfluid DM model, while \cite{Ho:2010ca, Ho:2011xc, Ho:2012ar} proposes 
a DM which is made aware of $\Lambda$ via gravitational thermodynamics \cite{Verlinde:2010hp}.
Various models that modify gravity have been considered in 
Refs.~\cite{Nojiri:2006su,Nojiri:2008nt,Brownstein:2005dr,Moffat:2013sja,Moffat:2013uaa,Moffat:2014bfa,Mendoza:2012qs,Bernal:2015fka,Barrientos:2020tcj,OBrien:2017bwr,Smolin:2017kkb,Verlinde:2016toy}.
However, in the case of $f(R)$ modified gravity models, it can be shown \cite{Nojiri:2006su,Nojiri:2008nt,Ketov:2018uel}
that the models are equivalent to usual gravity coupled in a non-standard way to some scalar field.
If this scalar field is interpreted as DM, then $f(R)$ models can also be considered a type of exotic DM theory.
Of course, this is to be expected since terms of the left-hand side of Eq.~\eqref{Einstein} can always be moved to the
right, and vice versa.
Both modified DM and modified gravity theories have a variety of phenomenological constraints that must be evaded, 
and it remains to be seem if any of these theories are successful in explaining BTFR and BFJR.

We conclude this letter by recalling that in particle physics, 
Bjorken scaling \cite{Bjorken:1967fb,Bjorken:1968dy} lead to the discoveries of asymptotic freedom, 
quarks, and eventually QCD.
We anticipate that the scalings we see in BTFR and BFJR will
also lead to important discoveries on DM.

\bigskip
\begin{acknowledgments}
We thank D.~Farrah, S.~Horiuchi, and J.~Y.~Ng for their collaboration on various related projects, and
L.~Freidel, C.~Frenk, J.~Khoury, R.~Leigh, P.~Mannheim, S.~McGaugh, J.~Moffat, and L.~Smolin for many helpful discusssions.
Preliminary versions of this work have been presented at MIAMI2018 and BASIC2019 \cite{Edmonds:2017fce}.
We thank E.~Guendelman and T.~Curtright for invitations to present at these conferences.
DM and TT are supported in part by the US Department of Energy (DE-SC0020262).
DM is also supported in part by the Julian Schwinger Foundation, 
and TT by the US National Science Foundation (NSF Grant 1413031).
\end{acknowledgments}


\bibliographystyle{apsrev4-2}
\bibliography{mdm}

\end{document}